\documentclass[10pt,prl,nofootinbib,notitlepage,twocolumn,superscriptaddress]{revtex4-1}

\usepackage[utf8]{inputenc}
\usepackage[T1]{fontenc}
\usepackage[english]{babel}  

\usepackage{times}

\usepackage{amssymb,amsmath,amsfonts,amsthm}
\usepackage{mathtools}
\usepackage{verbatim}
\usepackage{enumerate}
\usepackage{graphicx}
\usepackage{hyperref}
\usepackage{bbm}
\usepackage{booktabs}

\usepackage[caption=false]{subfig}
\graphicspath{{pics/}}
\usepackage{multirow}

\usepackage{color}
\definecolor{dred}{rgb}{.8,0.2,.2}
\definecolor{ddred}{rgb}{.8,0.5,.5}
\definecolor{dblue}{rgb}{.2,0.2,.8}
\definecolor{dgreen}{rgb}{.2,0.5,.2}

\newcommand{\bra}[1]{\mbox{$\langle #1|$}}
\newcommand{\ket}[1]{\ensuremath{|#1\rangle}}
\newcommand{\braket}[2]{\mbox{$\langle #1|#2\rangle$}}

\newcommand{\be}{\begin{equation}}
\newcommand{\ee}{\end{equation}}
\newcommand{\bea}{\begin{eqnarray}}
\newcommand{\eea}{\end{eqnarray}}

\newcommand{\tr}{\textrm{tr}}

\begin{document}

\preprint{APS/123-QED}

\title{Experimentally superposing two pure states with partial prior knowledge}

\author{Keren Li}
\affiliation{State Key Laboratory of Low-Dimensional Quantum Physics and Department of Physics, Tsinghua University, Beijing 100084, China}
\affiliation{Institute for Quantum Computing and Department of Physics and Astronomy, University of Waterloo, Waterloo N2L 3G1, Ontario, Canada}

\author{Guofei Long}
\affiliation{Institute for Quantum Computing and Department of Physics and Astronomy, University of Waterloo, Waterloo N2L 3G1, Ontario, Canada}

\author{Hemant Katiyar}
\affiliation{Institute for Quantum Computing and Department of Physics and Astronomy, University of Waterloo, Waterloo N2L 3G1, Ontario, Canada}

\author{Tao Xin}
\affiliation{State Key Laboratory of Low-Dimensional Quantum Physics and Department of Physics, Tsinghua University, Beijing 100084, China}

\author{Guanru Feng}
\affiliation{Institute for Quantum Computing and Department of Physics and Astronomy, University of Waterloo, Waterloo N2L 3G1, Ontario, Canada}

\author{Dawei Lu}
\email{d29lu@uwaterloo.ca}
\affiliation{Institute for Quantum Computing and Department of Physics and Astronomy, University of Waterloo, Waterloo N2L 3G1, Ontario, Canada}

\author{Raymond Laflamme}
\affiliation{Institute for Quantum Computing and Department of Physics and Astronomy, University of Waterloo, Waterloo N2L 3G1, Ontario, Canada}
\affiliation{Perimeter Institute for Theoretical Physics, Waterloo, Ontario, Canada}
\affiliation{Canadian Institute for Advanced Research, Toronto, Ontario M5G 1Z8, Canada}

\date{\today}

\begin{abstract}
Superposition, arguably the most fundamental property of quantum mechanics, lies at the heart of quantum information science. However, how to create the superposition of any two unknown pure states remains as a daunting challenge. Recently, it is proved that such a quantum protocol does not exist if the two input states are completely unknown, whereas a probabilistic protocol is still available with some prior knowledge about the input states [M. Oszmaniec \emph{et al.}, Phys. Rev. Lett. 116, 110403 (2016)]. The knowledge is that both of the two input states have nonzero overlaps with some given referential state. In this work, we experimentally realize the probabilistic protocol of superposing two pure states in a three-qubit nuclear magnetic resonance system. We demonstrate the feasibility of the protocol by preparing a families of input states, and the average fidelity between the  prepared state and expected superposition state is over 99\%. Moreover, we experimentally illustrate the limitation of the protocol that it is likely to fail or yields very low fidelity, if the nonzero overlaps are approaching zero. Our experimental implementation can be extended to more complex situations and other quantum systems.
\end{abstract}

\maketitle

\section{Introduction}
\label{intro}
Quantum superposition is a fundamental principle of quantum mechanics, and plays a crucial role in quantum information science \cite{Nielsen2010}. It states that, any two pure states can be added together or so-called superposed, while the outcome will be another valid pure state. As the central mystery of quantum physics, superposition leads to many unique quantum phenomena such as entanglement \cite{Horo2009} and interference \cite{Anderson1998, Zeil1999, Nobel2013}, and has many applications in quantum algorithms \cite{Shor1997,Grover1996,Grover1997} and quantum cryptography \cite{gisin2002quantum}. To further take advantage of this property,
one can ask if it is possible  to superpose two unknown pure states  generated by subroutines of a quantum algorithm in future quantum computers.

Such a task finally turns out to be impossible in a universal manner \cite{Alva2015, Horo2016}. Similar to the non-cloning theorem that an unknown state cannot be perfectly cloned without prior knowledge, the existence of a universal quantum protocol of superposing unknown pure states is also prohibited by the no-go theorem. In other words, it rules out the possibility of executing superposition operations for two completely unknown pure states. Nevertheless, a modified probabilistic protocol involving an ancilla qubit, a controlled-SWAP gate, and a post-selection process is shown in \cite{Horo2016}, if both of the two input states have fixed nonzero overlaps with a given referential state. This protocol can also be used to generate nonclassical states or non-Gaussian states efficiently in the context of quantum optics \cite{Horo2016}. Compared to the complete knowledge of the input states which usually requires full state tomography, the prior knowledge in this probabilistic protocol is much weaker and thus easy to be satisfied in a practical setup.

Very recently, the probabilistic scheme of superposing two pure states has been implemented using two photons \cite{hu2016}. Due to the lack of controllability in implementing the controlled-SWAP gate by weakly interacting photons, in that work the photonic system does not undergo the controlled-SWAP gate in a unitary manner, but by post-selecting some polarization subspace of the first photon. The polarization freedom on the first photon acts as a virtual ancilla qubit, and thus saves one qubit compared to the original protocol. In contrast to the photonic system, nuclear magnetic resonance (NMR) has been demonstrated to be a good testbed for quantum information tasks \cite{Ray12,Dawei14,Dawei15}, and its mature coherent control techniques \cite{PhysRevA.78.012328} enables the implementation of the controlled-SWAP gate with high fidelity.

In this work, we experimentally demonstrate the probabilistic protocol of creating a superposition of two unknown pure states in a three-qubit NMR system, where two qubits are used to generate two input states, and one qubit acts as the ancilla to carry out the controlled-SWAP gate. We implement two groups of experiments: fix the forms of input states and vary the superposition weights, and vice versa. Moreover, we discuss the efficiency of the protocol both in simulation and experiment, that is, when the fixed overlaps between the input states and the referential state are approaching zero, whether the protocol will succeed under realistic noisy environment. This problem is not accounted for in either of the previous works \cite{Horo2016, hu2016}, but worthwhile to be studied since the experimental imperfections are unavoidable.

The paper is organized as follows: In Sec. \ref{3bitscheme}, we briefly review the protocol and its circuit. In Sec. \ref{experiment}, we introduce our experimental setups, and the experimental procedure. In Sec. \ref{result}, we present the experimental results and discuss the drawbacks of the protocol in practice if the prior knowledge are poor. Finally, Sec. \ref{conclusion} summarizes the entire work and give some prospects in its future applications.

\section{theoretical overview}
\label{3bitscheme}
\begin{figure}[h]
    \centering\includegraphics[width=6cm]{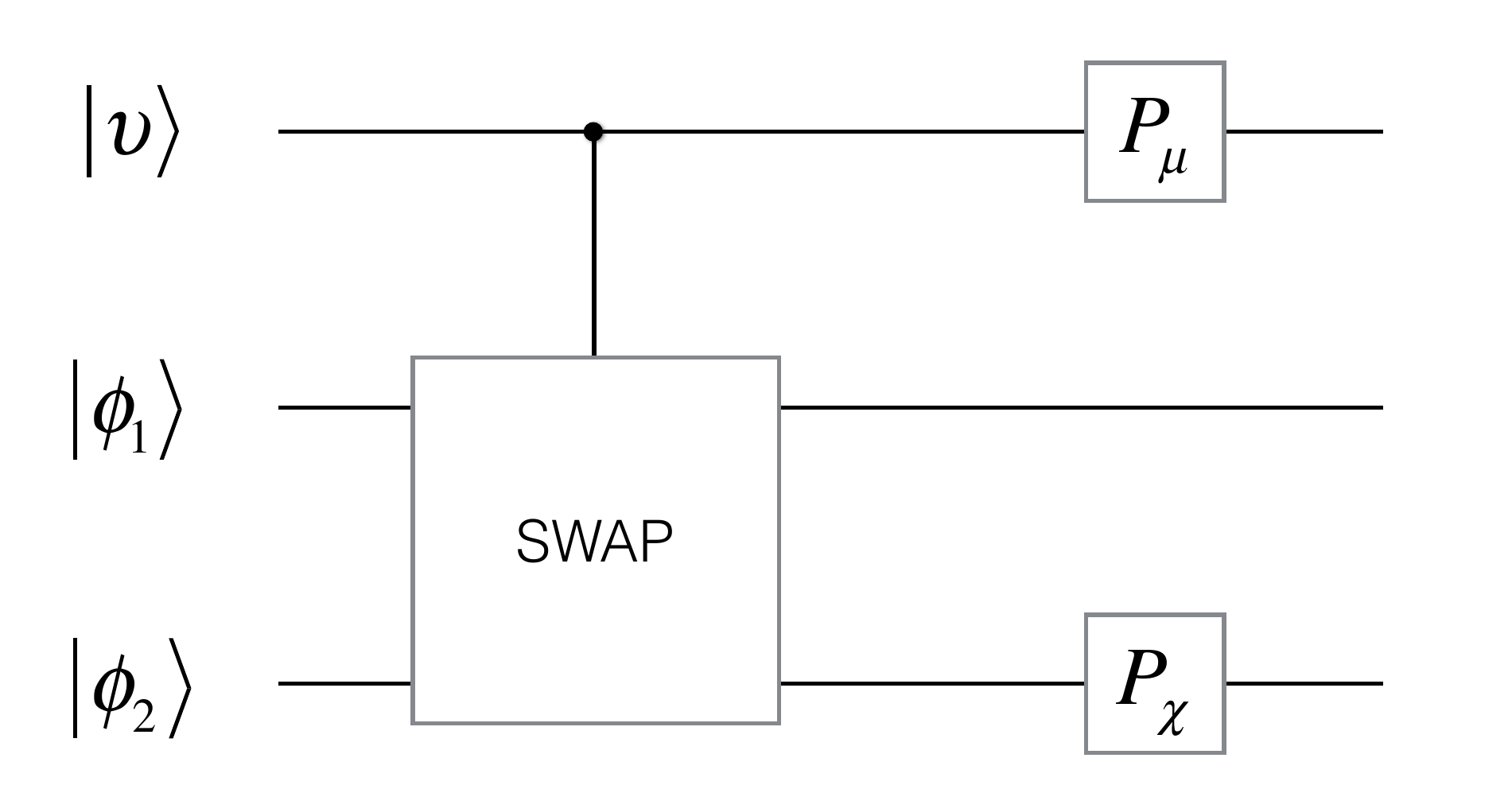}
    \caption{Probabilistic protocol of creating the superposition of two pure states.  $|\phi_{1}\rangle$, $|\phi_{2}\rangle$ are the two input states, and $|v\rangle$ is the ancilla state. Then the system undergoes a controlled-SWAP gate, which is the key ingredient of the scheme. $P_{\mu}$ is a projective measurement to project the ancilla qubit onto $|\mu\rangle$, and $P_{\chi}$ is to project the second input qubit(s) onto $|\chi\rangle$. $|\chi\rangle$  is the referential state chosen in the beginning, and $|\mu\rangle$ is computed by  $|\mu\rangle=|\langle\phi_{1}|\chi\rangle||0\rangle+|\langle\phi_{2}|\chi\rangle||1\rangle$, since we have already known the overlaps between the input states and referential state ($\langle\phi_{1}|\chi\rangle$ and $\langle\phi_{2}|\chi\rangle$).}
\label{scheme1}
\end{figure}

Whether there is any universal way to create superposition from two unknown pure states remained open, until a no-go theorem was proved by Alvarez-Rodriguez \emph{et al.} \cite{Alva2015} and Oszmaniec \emph{et al.} \cite{Horo2016} independently. In Ref. \cite{Alva2015}, it is proved that a quantum adder, which can be regarded as superposing two quantum states with the same weight, is forbidden in absence of an ancillary system. In Ref. \cite{Horo2016},  it is shown that creating the superposition of two pure states without any prior information is impossible. Subsequently,  Oszmaniec \emph{et al.} \cite{Horo2016} came up with an alternative protocol to produce the superposition from two input states, given that each of them has a fixed nonzero overlap with some referential pure state, as shown in Fig. \ref{scheme1}.

Assume that one would like to create the superposition $\ket{\phi_{\text{sup}}}$ of two states $\ket{\phi_1}$ and $\ket{\phi_2}$, such that
\be
\ket{\phi_{\text{sup}}}=\alpha|\phi_{1}\rangle+\beta|\phi_{2}\rangle.
\label{superform}
\ee
We then prepare the ancilla qubit into $|\nu\rangle=\alpha|0\rangle+\beta|1\rangle$, and drive the entire system to undergo a controlled-SWAP gate. When the ancilla is $\ket{0}$, the two input states swap their information, otherwise remain unchanged. Intuitively, this operation can be understood in the picture that the information of $\ket{\phi_2}$ is partially transferred to the Hilbert space of $\ket{\phi_1}$, depending on the coefficients $\alpha$ and $\beta$ in $|\nu\rangle$.

To obtain the superposition state, next we apply two projective measurements $P_{\mu}$ and $P_{\chi}$ on the ancilla and second input state, respectively. $|\chi\rangle$  is the referential state, and  $|\mu\rangle=|\langle\phi_{1}|\chi\rangle||0\rangle+|\langle\phi_{2}|\chi\rangle||1\rangle$. The form of $|\mu\rangle$ implies that we need to know the overlaps of $\langle\phi_{1}|\chi\rangle$ and $\langle\phi_{2}|\chi\rangle$ in advance, which is the sole condition in this protocol. After the projective measurements $P_{\mu}$ and $P_{\chi}$, the state in the space of $\ket{\phi_1}$ is the superposition state
\begin{equation}
\ket{\phi_{\text{sup}}}=\alpha\frac{\langle\chi|\phi_{2}\rangle}{|\langle\chi|\phi_{2}\rangle|}|\phi_{1}\rangle+\beta\frac{\langle\chi|\phi_{1}\rangle}{|\langle\chi|\phi_{1}\rangle|}|\phi_{2}\rangle.
\end{equation}
The whole procedure is depicted in Fig. \ref{scheme1}.

To demonstrate the above protocol in experiment, it is better to test its universality via two groups of experiments. For group A, we fix the input states and vary the coefficients $\alpha$ and $\beta$ in the superposition form in Eq. (\ref{superform}). The purpose is to show the protocol enables the superposition creation with arbitrary superposing weights. In experiment, the two input states are fixed as $|\pm\rangle$, and the variations of $\alpha$ and $\beta$ correspond to different states of ancilla that
\be
|\nu\rangle=\alpha|0\rangle+\beta|1\rangle = \text{cos}\frac{\theta_1}{2} \ket{0} + \text{sin}\frac{\theta_1}{2} \ket{1}.
\label{theta1}
\ee
For group B, we vary the input states and fix the coefficients, the purpose of which is to show that the protocol is able to perform a superposition operation for any two input states.  In experiment, we fix the ancilla state as $|\nu\rangle=|+\rangle$, and the first input state as $|0\rangle$, while altering the second input state via
\be
\ket{\phi_2}= \text{cos}\frac{\theta_2}{2} \ket{0} + i\text{sin}\frac{\theta_2}{2} \ket{1}.
\label{theta2}
\ee
Note that we add an imaginary unit before $\ket{1}$ to check the performance of the protocol when tackling states with phases.

 \begin{figure}[htpp]
    \centering\includegraphics[width=8cm]{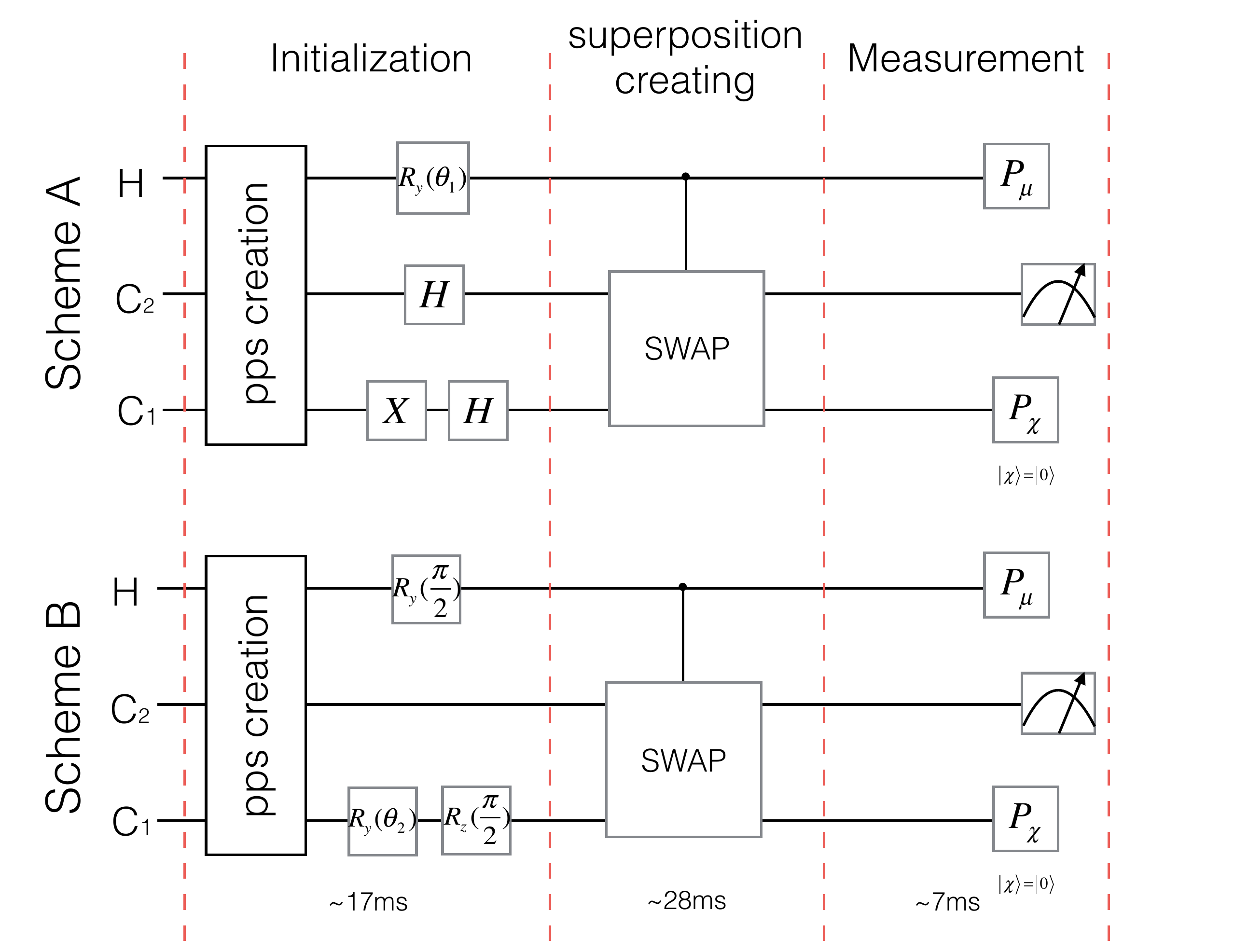}
    \caption{Implementation of the probabilistic protocol of superposing two pure states in a three-qubit NMR system. Group A and B experiments both consist of three steps, and the length of each step is shown at the bottom: the initialization is to create a three-qubit pseudo-pure state in about 15 ms, followed by carrying out 2 ms procedure to initialize the input states; the superposition creation is a controlled-SWAP gate realized in about 28 ms, after which the three-qubit system already contains the information of the superposition state on the second qubit; the measurement contains two projective measurements and one state tomography on the second qubit to extract the superposition state.  Despite lack of the genuine projective measurements in NMR, we mimicked them using gradient echo techniques \cite{PhysRevLett.86.5845}. The referential state is chosen as $\ket{\chi} = |0\rangle$. }\label{exp_cir}
\end{figure}

\section{experiment}
\label{experiment}
\begin{figure}[htpp]
    \centering\includegraphics[width=8cm]{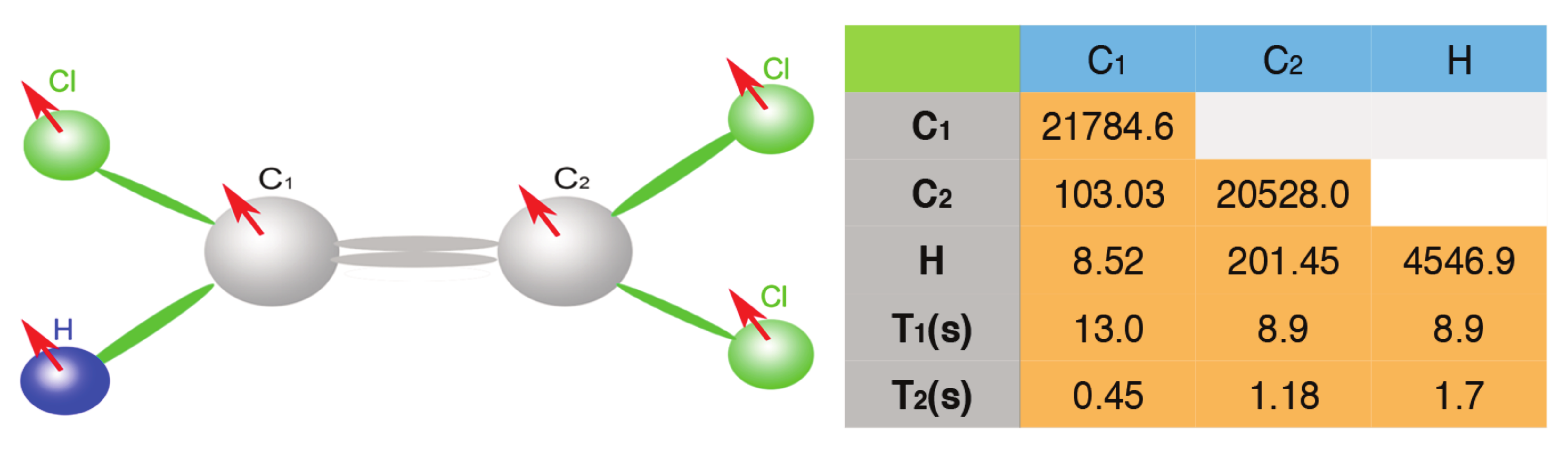}
    \caption{Molecular structure of TCE, where one $^{1}$H and two $^{13}$C nuclear spins form the three-qubit system. The table on the right lists the parameters of the internal Hamiltonian, in which the diagonal elements are chemical shifts (Hz), and the off-diagonal elements are the J-coupling strengths (Hz). T$_1$ and T$_2$ are the relaxation times (second) of the individual spins, respectively. All parameters are obtained on a Bruker DRX 700 MHz spectrometer at room temperature. }\label{tce}
\end{figure}

Now we turn to the description of our experimental setup.  All experiments were carried out on a Bruker DRX 700MHz spectrometer at room temperature. The three-qubit system is represented by $^{13}$C-labeled  trichloroethylene (TCE) molecule dissolved in d-chloroform (CDCl$_{3}$).  The TCE molecule, whose structure is shown in Fig. \ref{tce}, consists of two $^{13}$C's and one $^{1}$H. Here we use H as the ancilla qubit, C$_2$ as the qubit where $\ket{\phi_1}$ is encoded, and C$_1$ as the qubit where $\ket{\phi_2}$ is encoded, respectively. So the three qubits are assigned in the order of H, C$_2$ and C$_1$. As the frequency difference ($\sim$ 1250 Hz) between C$_1$  and C$_2$  is not sufficiently large compared to their J-coupling strength ($\sim$ 100 Hz), the coupling between C$_1$  and C$_2$ should be treated in the strong coupling regime. Hence, the internal Hamiltonian of the system is written as
\begin{eqnarray}\label{Hamiltonian}
\mathcal{H}=&&\sum\limits_{i=1}^3 {\pi \nu _i } \sigma _z^i  + \frac{\pi}{2}(J_{12}\sigma _z^1 \sigma _z^2+J_{13}\sigma _z^1 \sigma _z^3) \nonumber\\
&&+ \frac{\pi}{2}J_{23} (\sigma _x^2 \sigma _x^3+\sigma _y^2\sigma _y^3+\sigma _z^2 \sigma _z^3),
\end{eqnarray}
where $\nu_{i}$ is the chemical shift of the $i$th spin, $J_{ij}$ represents the J-coupling strength between the $i$th and $j$th spins. The values of all parameters are listed in the table of Fig. \ref{tce}.

\begin{figure}[h]
    \centering\includegraphics[width=8cm]{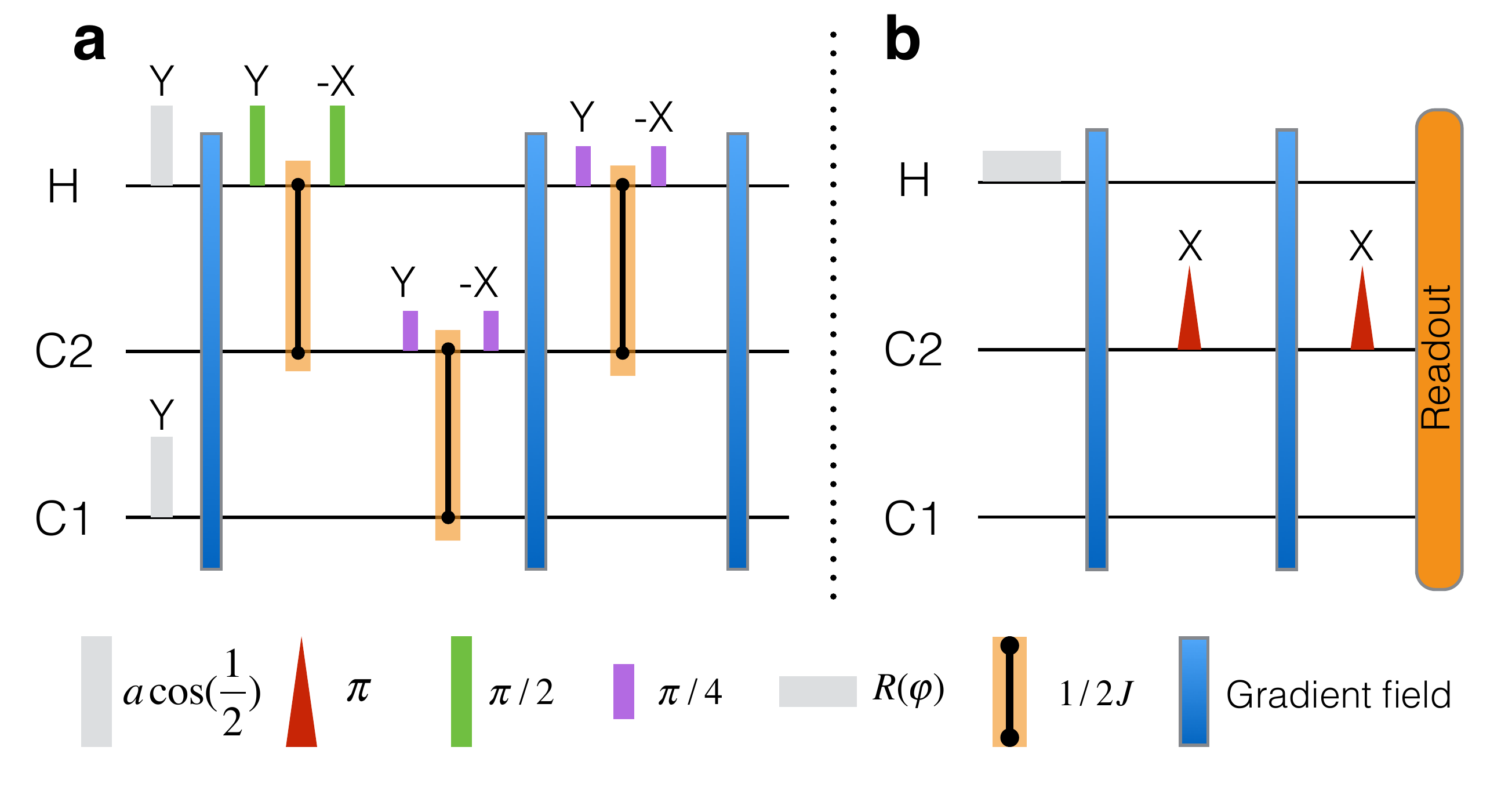}
    \caption{Experimental pulse sequences.  (a) Creating the pseudo-pure state via spatial average technique. The procedure includes local operations, three J-coupling evolutions, and three $z$-gradient pulses to destroy the unwanted coherent terms. The time of each free evolutions is $1/2J$ in the order of magnitude of 10 ms, determined by the J-coupling strengths of the relevant spins. (b) Simulating the projective measurements by gradient echoes. The procedure includes two $z$-gradient pulses and two $\pi$ pulses. $R(\mu)$ is a single-qubit pulse to rotate $|0\rangle$ to $|\mu\rangle$.}\label{pps_cir}
\end{figure}

The experiment started from the thermal equilibrium state
\begin{equation}
\rho_{eq}=\frac{1-\epsilon}{8} \mathbb{I} + \epsilon (\gamma_{\text{H}} \sigma_{z}^{1}+\gamma_{\text{C}} \sigma_{z}^{2}+\gamma_{\text{C}} \sigma_{z}^{3}),
\end{equation}
where $\epsilon \sim 10^{-5}$ describes the polarization of the entire system, $ \mathbb{I}$ is a $8\times 8$ identity matrix, and $\gamma_{\text{H}}$ and $\gamma_{\text{C}}$ are the gyromagnetic ratios of $^{1}H$ and $^{13}C$, respectively. As the identity matrix does not evolve under unitary operations, and cannot be detected in the spectrometer, we simply neglect it and rewrite $\rho_{eq}$ as
\begin{equation}
\label{drho}
\rho_{eq}^{dev} \approx 4 \sigma_{z}^{1} + \sigma_{z}^{2} + \sigma_{z}^{3},
\end{equation}
where we use the fact that $\gamma_{\text{H}}\approx 4\gamma_{\text{C}}$.
The form of Eq. (\ref{drho}) is called deviation density matrix, and is often used to replace the original density matrix which contains very large identity in NMR. From the deviation density matrix, we employed the spatial average technique \cite{Cory04031997, PhysRevLett.116.230501} as shown in Fig. \ref{pps_cir}(a) to create the pseudo-pure state (PPS). The final state is $\ket{000}$, where the identity term has been neglected.
In experiment, we realized the unitary operations in between the $z$-gradient pulses (see Fig. \ref{pps_cir}(a)) via three shaped pulses optimized by the gradient ascent pulse engineering (GRAPE) technique \cite{khaneja2005}. The total length of the PPS creation is about 15 ms with the simulated fidelity 99.8\%, while in experiment we reached the fidelity over 99\%.

Next we prepared the system into the initial states in group A and B, respectively. The two preparations circuits are displayed in Fig. \ref{exp_cir}. In group A, we varied $\theta_1$ in Eq. (\ref{theta1}) by 12 values from $0$ to $\pi$, where each $\theta_1$ indicated different superposing weights for $\ket{\phi_1}$ and $\ket{\phi_2}$. In group B, we varied $\theta_2$ in Eq. (\ref{theta2}) by 12 values from $0$ to $\pi$, where each $\theta_2$ stood for a different input state $\ket{\phi_2} = \text{cos}\frac{\theta_2}{2} \ket{0} + i\text{sin}\frac{\theta_2}{2} \ket{1}$. All pulses during the preparation stage were local and optimized by 2 ms GRAPE pulses with the simulated fidelities over 99.9\%.

After the initializing step, we would create the superposition of the prepared $\ket{\phi_1}$ and $\ket{\phi_2}$, where the superposing weights were determined according to the state of the ancilla qubit $\ket{\nu}$. The superposition operation itself can be accomplished via a controlled-SWAP gate, where H is the control qubit as shown in Fig. \ref{exp_cir}. In experiment, this complicated gate was optimized by a 28 ms GRAPE pulse with fidelity of 99.9\%, as well as robustness to the drift of the internal Hamiltonian and inhomogeneity of the control field. Whereafter, this pulse was rectified via a feedback control device attached to the spectrometer to minimize the discrepancies between the calculated pulse and its experimental performance. Compared to the traditional circuit decomposition approach, the utilization of GRAPE technique and feedback control technique for the implementation of the controlled-SWAP gate does boost our experimental results greatly, as we will see later.

The last step in the original scheme requires two projective measurements. The main purpose is to collapse the whole system onto a one-qubit subspace, where the residual state will be the expected superposition. Projective measurements can be mimicked non-selectively in NMR using a combination of pulsed gradient fields and some refocusing pulses \cite{PhysRevLett.86.5845,xiao2006nmr,auccaise2012experimental}. In this work, we implement this step by inserting a pair of gradient pulses and refocusing pulses, as shown in Fig.\ref{pps_cir}(b), where the entire procedure is about 7ms.

To summarize our experimental procedure, we first created the PPS with over 0.99 fidelity. We then varied $\theta_1$ in Eq. (\ref{theta1}) from 0 to $\pi$ for experiments in group A, and $\theta_2$ in Eq. (\ref{theta2}) for group B. Next, a high-fidelity 28 ms GRAPE pulse was applied to achieve the controlled-swap gate, followed by a 7 ms gradient echo sequence to mimic the projective measurements.

\section{Experiment Result}
\label{result}

\begin{table*}[htpp]
\caption{All fidelities between the theoretical superposition states and experimentally prepared states. The upper part shows the results of group A, where the two input states are fixed and ancilla is varied; the lower part shows the results of group B, where the second input state is  varied and all the other two are fixed. $\theta_1$ and $\theta_2$ are defined in Eq. (\ref{theta1}) and  Eq. (\ref{theta2}), respectively.}
\begin{center}
\begin{tabular}{|c|c|c|c|c|c|c|c|}
\hline
\multirow{4}{*}{A}&$\theta_1$ & 0 & $\pi$/12 & 2$\pi$/12 & 3$\pi$/12 & 4$\pi$/12 & 5$\pi$/12 \\
\cline{2-8}
&fidelity & 0.9875$\pm$0.0020 & 0.9902$\pm$0.0023 & 0.9838$\pm$0.0020 & 0.9750$\pm$0.0033 & 0.9782$\pm$0.0022 & 0.9830$\pm$0.0026 \\
\cline{2-8}
&$\theta_1$ & 6$\pi$/12 & 7$\pi$/12 & 8$\pi$/12 & 9$\pi$/12 & 10$\pi$/12 & 11$\pi$/12 \\
\cline{2-8}
&fidelity & 0.9900$\pm$0.0026 & 0.9901$\pm$0.0028 & 0.9911$\pm$0.0030 & 0.9897$\pm$0.0067 & 0.9857$\pm$0.0012 & 0.9930$\pm$0.0014 \\
\hline
\hline
\multirow{4}{*}{B}&$\theta_2$ & 0 & $\pi$/12 & 2$\pi$/12 & 3$\pi$/12 & 4$\pi$/12 & 5$\pi$/12 \\
\cline{2-8}
&fidelity & 0.9948$\pm$0.0018 & 0.9882$\pm$0.0055 & 0.9920$\pm$0.0080 & 0.9800$\pm$0.0072 & 0.9918$\pm$0.0026 & 0.9924$\pm$0.0060 \\
\cline{2-8}
&$\theta_2$ & 6$\pi$/12 & 7$\pi$/12 & 8$\pi$/12 & 9$\pi$/12 & 10$\pi$/12 & 11$\pi$/12 \\
\cline{2-8}
&fidelity & 0.9880$\pm$0.0087& 0.9802$\pm$0.0119 & 0.8968$\pm$0.0285 & 0.8512$\pm$0.0389 & 0.7295$\pm$0.0451 & 0.6867$\pm$0.0739\\
\hline
\end{tabular}
\end{center}
\label{fid}
\end{table*}%

For group A where the two input states are fixed and ancilla is varied, we computed the fidelity between the theoretical superposition state $\rho_{\text{th}} = \ket{\phi_{\text{sup}}}\bra{\phi_{\text{sup}}}$, where $\ket{\phi_{\text{sup}}}$ is shown in Eq. (\ref{superform}), and the experimentally created superposition state $\rho_{\text{exp}}$ via
\begin{equation}
F=\tr(\rho_{\text{th}}\rho_{\text{exp}})/\sqrt{\tr(\rho_{\text{th}}^{2})\tr(\rho_{\text{exp}}^{2})}.
\end{equation}
Here, we write the states in terms of density matrices because the experimentally prepared state is mixed due to the experimental imperfections. All 12 fidelities in group A are shown in the upper part of Table \ref{fid}, and it can be seen that each fidelity is around 0.98, meaning that the probabilistic protocol is valid under realistic noises.

\begin{figure}[h]
    \centering\includegraphics[width=0.85\columnwidth]{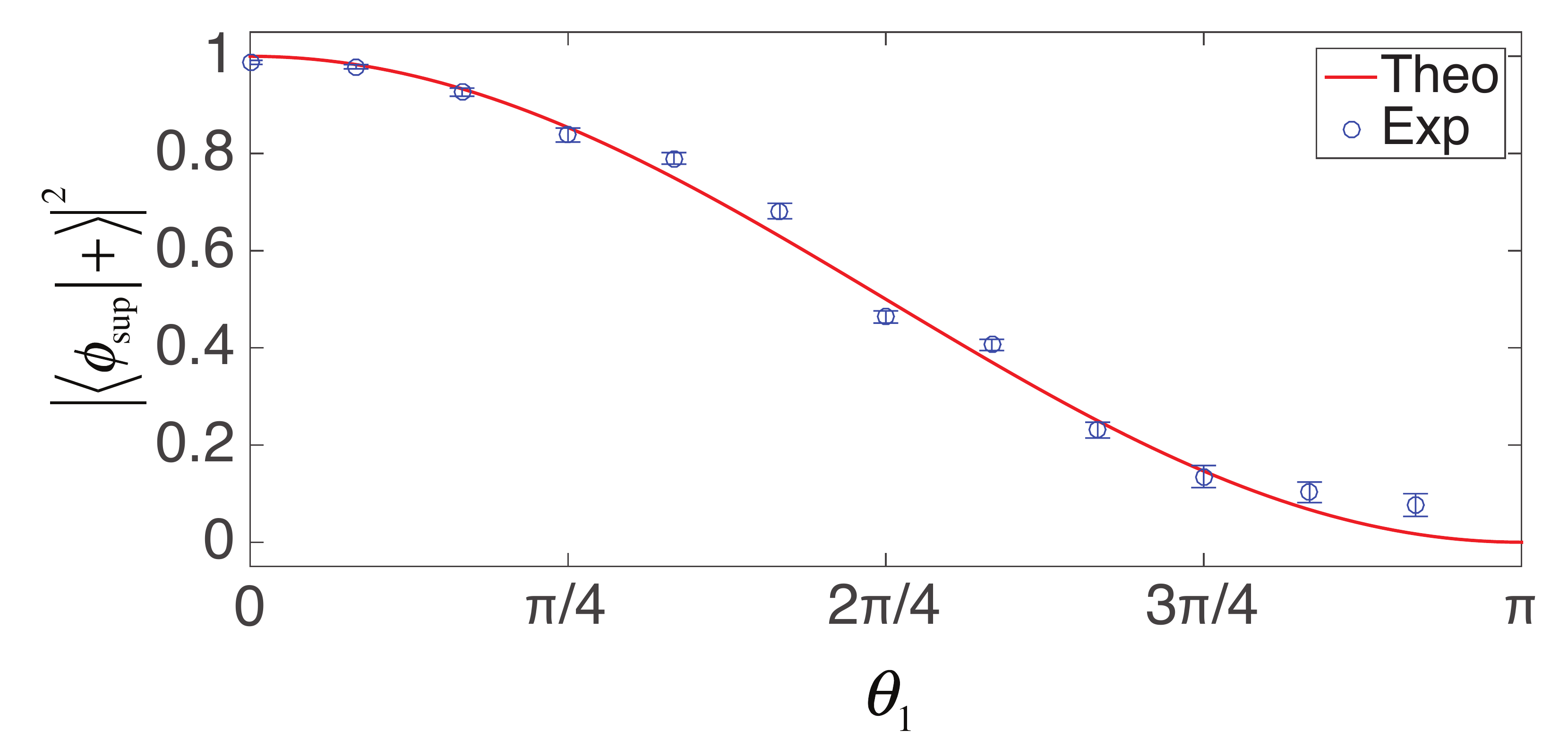}
    \caption{Overlaps between the final superposition state $\ket{\phi_\text{sup}}$ and the first input state $\ket{+}$, defined by $|\braket{\phi_\text{sup}} +|^2$. The red and blue are theoretical and experimental results, respectively. }\label{s1}
\end{figure}

Moreover, we exhibit the group A results in another way via the overlap $|\braket{\phi_\text{sup}} {\phi_1}|^2$ between the final superposition state $\ket{\phi_\text{sup}}$ and the first input state $\ket{\phi_1}= \ket{+}$, to show that the superposing weight of $\ket{\phi_1}$ is changing as we expected. The three-qubit state before the controlled-SWAP gate is $(\cos\frac{\theta_1}{2}|0\rangle+\sin\frac{\theta_1}{2}|1\rangle)\otimes|+\rangle\otimes|-\rangle$, and the outcome after the entire circuit is $\ket{\phi_\text{sup}} = \cos\frac{\theta_1}{2}|+\rangle+\sin\frac{\theta_1}{2}|-\rangle$. Thus, the theoretical value of the overlap is just
\begin{equation}
\mathcal{P}_{\text{A}}=\cos^{2}\frac{\theta_1}{2}.
\label{p1}
\end{equation}
The red curve in Fig. \ref{s1} represents the theoretical values of the overlap, while the blue bars are the experimental results with error bars. The error bars come from the uncertainty during the spectrum read out stage, which are the fitting errors in extracting the density matrices from the spectra explicitly.

For group B where the second input state is  varied and the first input and ancilla are fixed, all 12 fidelities are shown in the lower part of Table \ref{fid}. Compared to the group A result where all fidelities are around 0.99, the result in group B has the trend that the fidelity and uncertainty shown in Table \ref{fid} both go worse when $\theta_2$ approaches $\pi$. In other words, when the prior knowledge $\langle\phi_{2}|\chi\rangle$ approaches zero, the protocol is likely to fail or leads to bad fidelities.

\begin{figure}[h]
    \centering\includegraphics[width=0.85\columnwidth]{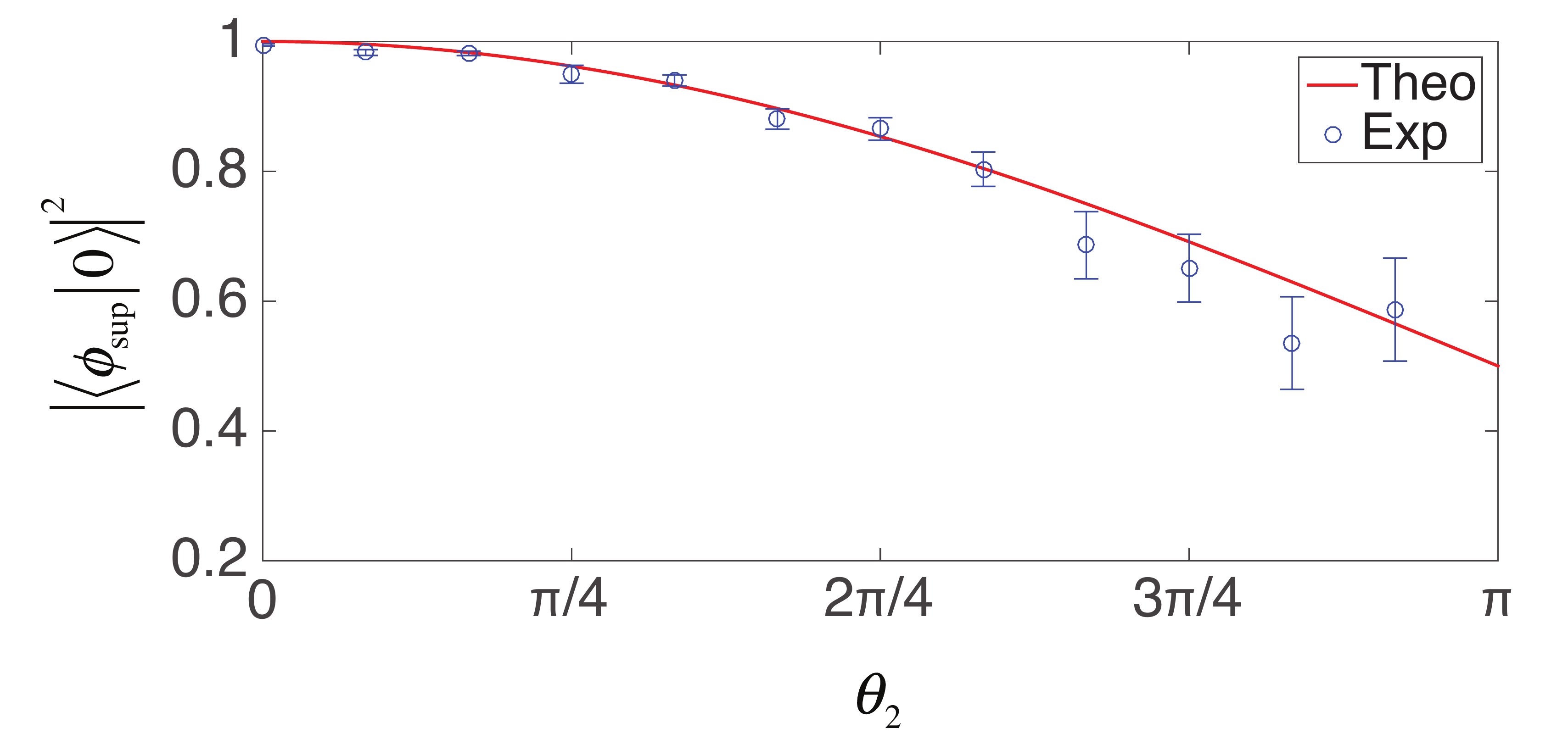}
    \caption{Overlaps between the final superposition state $\ket{\phi_\text{sup}}$ and the first input state $\ket{0}$, defined by $|\braket{\phi_\text{sup}} 0|^2$. The red and blue are theoretical and experimental results, respectively.}\label{s2}
\end{figure}

Similarly to the case of group A, we also consider the overlap $|\braket{\phi_\text{sup}} {\phi_1}|^2$ between the final superposition state $\ket{\phi_\text{sup}}$ and the first input state $\ket{\phi_1}=\ket{0}$. Before the controlled-SWAP gate, the three-qubit state is $\frac{\sqrt{2}}{2}(|0\rangle+ |1\rangle)\otimes|0\rangle\otimes (\cos\frac{\theta_2}{2}|0\rangle+\sin\frac{\theta_2}{2}|1\rangle)$. At the end of the circuit, the outcome of the superposition state is
\be
\ket{\phi_\text{sup}} = \frac{(1+\cos\frac{\theta_2}{2})|0\rangle+i\sin\frac{\theta_2}{2}|1\rangle}{\sqrt{2+2\cos\frac{\theta_2}{2}}}.
\ee
So the overlap between $\ket{\phi_\text{sup}}$ and $\ket{0}$ is
\begin{equation}
 \mathcal{P}_{\text{B}}=\frac{1+2\cos(\frac{\theta_2}{2})+\cos^{2}(\frac{\theta_2}{2})}{2+2\cos(\frac{\theta_2}{2})}.
 \label{p2}
\end{equation}
In Fig. \ref{s2}, we see that the uncertainty basically goes up along with the increase of $\theta_2$, which is consistent with the result in Table \ref{fid}.

Now we discuss the origin of the bad fidelities and uncertainties when the overlap $\langle\phi_{2}|\chi\rangle$ approaches zero. From the circuit in Fig. \ref{exp_cir}, we know that all other units do not directly depend on this overlap value except $P_{\mu}$, so the invalidity of the protocol mainly comes from $P_{\mu}$ in experiment, that is measurement section. Theoretically, the protocol is always valid as long as neither of the overlaps $\langle\phi_{1}|\chi\rangle$ and $\langle\phi_{2}|\chi\rangle$ is zero, no matter how small they are. However, in practice, a small number usually means it is more sensitive to the noise, especially when it appears in a denominator. In order to support our statement, we numerically simulated the changes of uncertainty for different values of $\langle\phi_{1}|\chi\rangle$ and $\langle\phi_{2}|\chi\rangle$. In our simulation, we take into account the following errors: state preparation, that is, our simulation started from the experimental PPS reconstructed by tomography but not the theoretical PPS; unitary (coherent) errors in the evolution, as our GRAPE pulses in implementing the circuits are not with fidelity one; decoherence (incoherent) errors in the evolution, using the relaxation parameters in Fig. \ref{tce} and the procedure in \cite{Dawei15}; measurement errors including the gradient echo and readout. In other words, this is a realistic simulation of the experimental environment  to our best knowledge.

\begin{figure}[h]
    \centering\includegraphics[width=0.9\columnwidth]{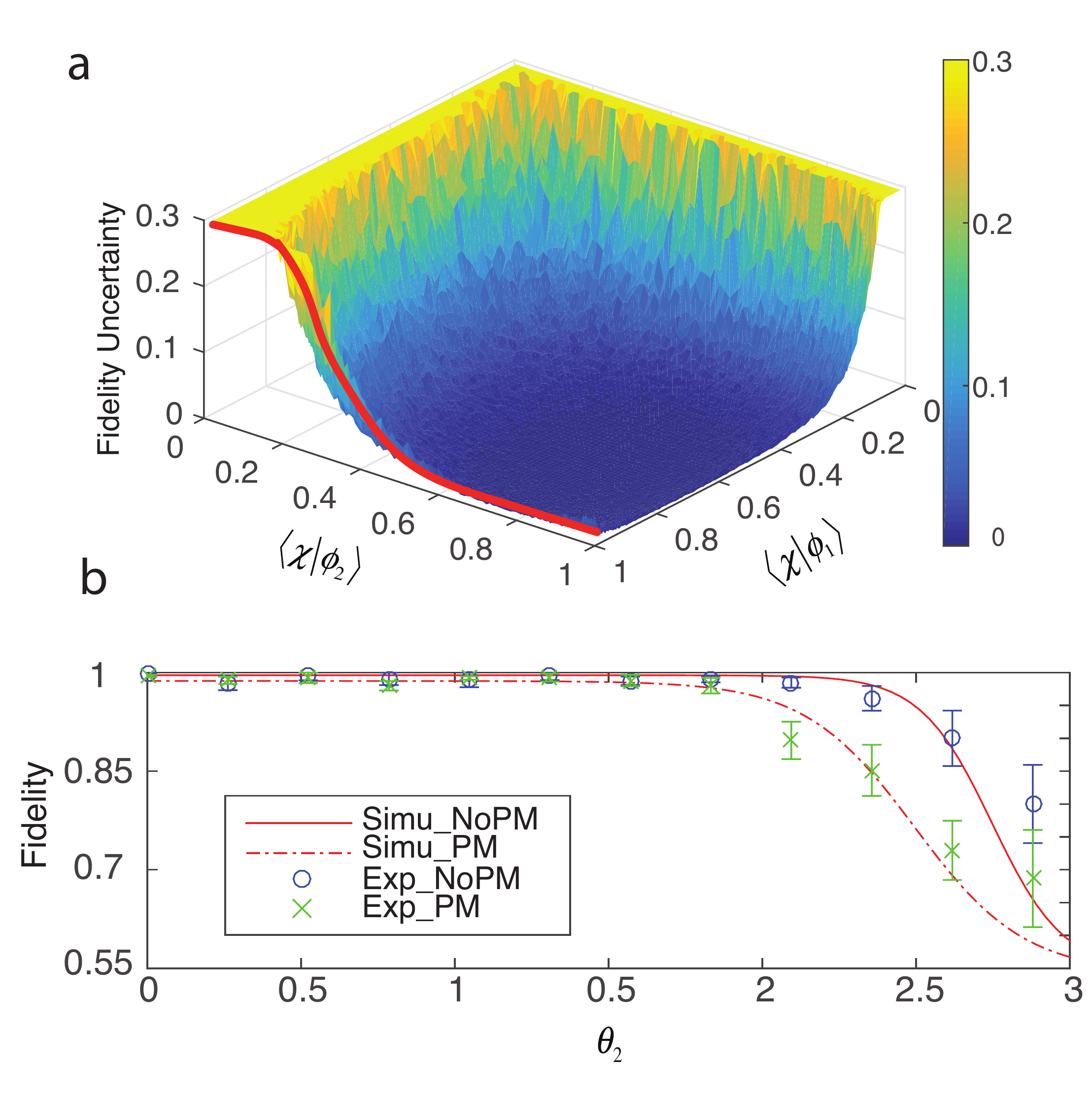}
    \caption{(a) Fidelity uncertainty for different values of overlaps $\langle\phi_{1}|\chi\rangle$ and $\langle\phi_{2}|\chi\rangle$, where $|\phi_{1}\rangle$ and $|\phi_{2}\rangle$ are the input states, and $|\chi\rangle$ is the referential state. Basically, the smaller the overlap is, the larger the uncertainty suffers. The red curve is the case for our group B experiments. (b) Fidelities with and without the implementation of projective measurements for different values of $\theta_2$. Notice that the overlap $\langle\phi_{2}|\chi\rangle = \cos(\theta_{2}/2)$. We see that mimicking the projective measurements induces extra errors and uncertainties in experiment.}\label{simu}
\end{figure}

The simulated result is shown in Fig. \ref{simu}(a). It is clear that a small value of overlap is responsible for the large error bar in experiment. In Fig. \ref{simu}(a), when either of the two overlaps falls down below some threshold value, the created superposition becomes unstable due to its high uncertainty. This simulation is able to explain the outliers in experiment, where the red curve in Fig. \ref{simu}(a) is actually the case we used for group B. Moreover, we experimentally test how the gradient echo, i.e. implementation of projective measurements, impacts the large overlap requirement by repeating group B experiments twice with and without the echo. For the case without the gradient echo, we directly carry out a three-qubit state tomography after the controlled-SWAP gate and trace out the unwanted two qubits numerically.  The simulated and experimental results are shown in Fig. \ref{simu}(b), where we can see that the implementation of projective measurements indeed causes extra errors. Therefore, to maintain a higher accuracy, a larger overlap is needed to tolerate the errors in the measurement process. It is worthy stressing that, the simulations in Fig. \ref{simu} are based on our experimental environment, so they may differ in other systems with different noise models.  Nevertheless, we can conclude that when the experimental noise gets stronger, the required overlaps $\langle\phi_{1}|\chi\rangle$ and $\langle\phi_{2}|\chi\rangle$ need to be larger in order to implement the protocol with high confidence.

\section{conclusion}
\label{conclusion}
Superposing two unknown pure states is a useful and interesting topic. It can be regarded as a quantum adder on vectors in Hilbert space, which may be embedded as a basic unit in future quantum computers to produce the final result of independent runs of quantum algorithms. For example, one important task in machine learning is to guess the output from a new input based on provided data points, and a quantum linear regression algorithm can be applied to solve it \cite{schuld2016}. The initial step of this algorithm is to prepare two states entangled with an ancilla qubit, which is straightforward to accomplish by superposing those two states coherently.

 Despite the lack of a universal way for implementation \cite{Alva2015, Horo2016}, a probabilistic protocol  is still available provided some prior knowledge about the input states \cite{Horo2016}. Very recently, the experimental demonstration of the validity of the protocol has been finished by photons, but without realizing the controlled-SWAP gate in a unitary manner \cite{hu2016}.  In this work, we show the feasibility of the protocol in a more general setting with a real ancilla qubit and unitarily implementing the controlled-SWAP gate.

Our experiment involves a three-qubit system, and demonstrates that the superposing weights are totally tunable as long as the prior knowledge about the overlaps are large enough. In contrast, when either of the two overlaps between the input states and referential state approaches zero, the entire protocol is likely to fail since the fidelity and uncertainty are both bad in this situation. Our simulation of the realistic noise model is consistent with the experimental result, which supports our statement that the prior knowledge should be sufficiently good to run the conditional quantum adder smoothly in practice.

 As a valid and proof-of-principle demonstration of creating a superposition of two unknown pure states, our experiments have effectively verified probabilistic protocol, and are beneficial to the emergence of, for instance, a practical quantum adder.

\begin{acknowledgements}
We are grateful to the following funding sources: Industry Canada, NSERC and CIFAR (H.K., G.F., D.L., and R.L.); National Natural Science Foundation of China under Grants No. 11175094 and No. 91221205 (K.L., and T.X.); National Basic Research Program of China under Grant No. 2015CB921002 (K.L., and T.X.); China Scholarship Council (K.L.).
\end{acknowledgements}



\end{document}